\documentclass{openjournal}

\usepackage{amsmath}
\usepackage{multirow}

\usepackage{xcolor}
\usepackage{textgreek}
\usepackage[utf8]{inputenc}
\usepackage[english]{babel}

\usepackage{hyperref}
\hypersetup{
    unicode, 
    colorlinks=true,
    linkcolor=linkcolor,
    citecolor=linkcolor,
    filecolor=linkcolor,
    urlcolor=linkcolor,
}
\usepackage{color,colortbl}
\definecolor{linkcolor}{rgb}{0.0,0.3,0.5}
\usepackage{tensind}
\tensordelimiter{?}
\DeclareGraphicsExtensions{.bmp,.png,.jpg,.pdf}
\usepackage{verbatim}
\usepackage[normalem]{ulem}
\usepackage{orcidlink}
\usepackage{soul}

\graphicspath{ {./figs/} }

\begin{document}
\title{The matter with(in) CPL}

\author{Leonardo Giani \orcidlink{orcid=0000-0001-6778-1030}}
\email{uqlgiani@uq.edu.au}
\affiliation{School of Mathematics and Physics, The University of Queensland,
 Brisbane, QLD 4072, Australia}

\author{Rodrigo von Marttens\orcidlink{0000-0000-0000-0000}}
\affiliation{Instituto de Fisica, Universidade Federal da Bahia, 40210-340, Salvador, BA, Brazil}
\affiliation{PPGCosmo, Universidade Federal do Espírito Santo, Vitória-ES, 29075-910, Brazil}

\author{Oliver Fabio Piattella\orcidlink{0000-0000-0000-0000}}
\affiliation{Dipartimento di Scienza e Alta Tecnologia, Universita' degli Studi dell’Insubria, via Valleggio 11, I-22100 Como, Italy}
\affiliation{INFN Sez. di Milano, Via Celoria 16, 20126, Milano, Italy.}
\affiliation{Nucleo Cosmo-ufes, Universidade Federal do Espirito Santo, avenida F. Ferrari 514, 29075-910 Vitoria,
Espirito Santo, Brazil.}



\begin{abstract}
We introduce a two-parameter phenomenological extension of the $\Lambda$CDM model in which the equation of state parameter of the ``dust'' fluid becomes different from zero for redshifts below a transition value $z_t$. 
Using data from DESI DR2 BAO, DESY5 Sn~Ia and CMB distance priors ($R,l_A,\omega_b$) data, we compare our model with the standard CPL parameterization $w_0-w_a$ for dynamical dark energy. Using the Deviance Information Criteria (DIC), we find that the two models are essentially indistinguishable ($\Delta$DIC $<$ 2) and preferred over $\Lambda$CDM with a significance $\geq 3 \sigma$. We discuss how this parameterization finds a natural interpretation in the context of cosmological backreaction and derive a prediction for the evolution of the growth factor, discussing its impact on low redshift $f\sigma_8$ measurements.    

\end{abstract}


\section{Motivation} 
During the past couple of years, the combination of Sn~Ia and BAO measurements from DES \citep{DES:2024tys} and DESI \citep{DESI:2024mwx,DESI:2024hhd,DESI:2025zgx} with CMB observations \citep{Planck:2018vyg} (or weak lensing measurements \citep{DES:2021wwk}) revealed a $\sim 3 \sigma$'s strong preference for Dynamical Dark Energy (DDE) over the cosmological constant $\Lambda$. This preference is usually quantified using the Chevallier-Polarsky-Linder (CPL) parametrization \citep{Chevallier:2000qy,Linder:2002et}, essentially a first order Taylor expansion at low redshift of the dark energy equation of state parameter $w$, but it has been shown to be robust even considering different dark energy models or parametrizations \citep{Giare:2024gpk, Jiang:2024xnu,Roy:2024kni,Sakr:2025daj,Wolf:2025jlc}. Whilst this exciting result might indeed provide insights on the fundamental nature of Dark Energy, the fact that its equation of state parameter crosses the ridge $w=-1$ is concerning from a theoretical point of view, as it suggests a dark energy fluid exhibiting phantom behavior. 
For these reasons, a number of works \citep{Cortes:2024lgw,Shlivko:2024llw,Ye:2025ark,RoyChoudhury:2024wri,Colgain:2024mtg,Gialamas:2024lyw,Colgain:2024xqj,Wolf:2024stt,Colgain:2025nzf,Wolf:2025jed,RoyChoudhury:2025dhe} have tried to understand what feature of the data drives the preference for $w_0w_a$ over the cosmological constant and whether or not phantom crossing is a necessary condition or just an artifact of the chosen parametrization. The individual datasets, when interpreted using a flat $\Lambda$CDM model, somewhat disagree on their inferred value of the present day matter density. In fact, BAO prefers $\Omega_m^{\rm{DESI}} \approx 0.29 \pm 10^{-3}$, whereas CMB indicates $\Omega_m^{\rm{Planck}} \approx 0.31 \pm 10^{-3}$ and supernovae $\Omega_m^{\rm{DESY5}} \approx 0.35 \pm 10^{-2}$. One would naturally conclude that $\Lambda$CDM seemingly does not have sufficient freedom (or parameters) to provide a consistent cosmological evolution.

With the above motivation in mind, in this work we introduce a novel two-parameter extension of the $\Lambda$CDM model, where rather than modifying the evolution of the dark energy fluid, we focus on the matter equation of state parameter. We formulate the following ansatz for the total (baryonic plus dark ) matter density:
\begin{equation}\label{non-dust}
    \rho_m = \tilde{\rho}^{\rm{dust}}_0\left(1+z\right)^3 f(z_t,\epsilon,z)\;, \qquad f(z_t,\epsilon,z)\equiv \left[ \Theta \left(z-z_t\right) + \frac{\left(1+z \right)^\epsilon}{\left(1+z_t\right)^\epsilon}\Theta \left(z_t-z\right) \right]\;,
\end{equation}
describing a perfect fluid whose equation of state parameter transitions from $w=0$ for $z>z_t$ to $w=\frac{\epsilon}{3}$ at later times $z<z_t$. In the above, $\Theta(x)$ is the Heaviside theta function: $\Theta(x) = 1$ if $x>0$ and $\Theta(x) = 0$ otherwise, and where $\tilde{\rho}^{\rm{dust}}_0$ represents the density that a standard dust fluid (without transition) would have at redshift zero starting from the same initial conditions. One can easily compute the difference between the two at present day:
\begin{equation}\label{present-day-matter}
    \rho_m(z=0)=\frac{\tilde{\rho}^{\rm{dust}}_0}{\left(1+z_t\right)^\epsilon}\;.
\end{equation}

This parametrization finds a natural interpretation in the context of cosmological backreaction \citep{Buchert:1995fz,Clarkson:2011zq,Bolejko:2016qku}. The distribution of the matter ``fluid elements'' across cosmic epochs is very different. Early in the evolution of the Universe, perturbations of the matter fluid are small, as reflected by the temperature fluctuations of the CMB. However, they are bounded to grow in an expanding Universe, eventually resulting in the complex web of inhomogeneities that we observe in the Universe today. The possible impact of these inhomogeneities on cosmological measurements and the overall expansion history of the Universe is subject of debate, with a spectrum of possibilities ranging from negligible impact \citep{Kaiser:2017hqn,Ishibashi:2005sj,Green:2010qy} to being responsible for the observed accelerated expansion of the Universe \citep{Rasanen:2003fy,Wiltshire:2009db,Buchert:2007ik}. It has also been recognized that in this era of high-precision cosmology, at least in our cosmic neighborhood, these inhomogeneities can lead to systematic biases in our cosmological inference even in the framework of the $\Lambda$CDM model \citep{Heinesen:2020pms,Bolejko:2015gmk,Giani:2023aor,Giani:2024nnv,Heinesen:2021azp,Anton:2023icm}. 
For our purposes, it is enough to recognize that the strong gravitational interactions between dark matter fluid elements in the late Universe at various scales justify possible deviations on the overall equation of state parameter of the matter fluid averaged on large volumes, as described by Eq.~\eqref{non-dust}. We model the transition to such a different behavior with an instantaneous process (as reflected by the Heaviside step distribution), because properly characterizing a smoothed process would unavoidably introduce a new parameter $\delta \tau$ accounting for the transition timescale, and as thoroughly discussed in Refs.~\citep{Linder:2005ne,DiValentino:2017zyq}, current observations are insufficient to effectively constrain dark energy parametrizations with more than two free parameters. In summary, Eq. \eqref{non-dust} can be considered as an effective description of a matter fluid whose averaged (integrated) equation of state parameter between $0<z<z_t$ is $\epsilon$.  
Volume averages of cosmological quantities in inhomogeneous spacetimes, as described for example in \cite{Buchert:2002ij,Buchert:2011sx}, result in modified Friedmann equations containing contributions from the kinematic backreaction and the spatial curvature terms, which in the past have been suggested as a possible alternative to dark energy. In this work, however, we rather assume that for an observer with no information about the true geometry of the spacetime, these terms are degenerate with what they identify as "matter", leading to a different cosmological evolution for the latter. This applies both to the case of local inhomogeneities \citep{Marra:2022ixf,ChirinosIsidro:2016tqo,Giani:2023aor}, such as an observer within an LTB region, and to the case of average over multiple inhomogeneities, such as the models discussed in \cite{Giani:2024nnv,Marra:2007pm,Barbosa:2015wlv}.

For the remainder of this letter we will explore the impact of our modeling choice on the expansion history, constraining the parameters $\epsilon$ and $z_t$ and performing a Bayesian comparison of the model against the $w_0 w_a$ parametrization and $\Lambda$CDM.

\begin{figure}[h!]
    \centering
    \includegraphics[width=0.5\linewidth]{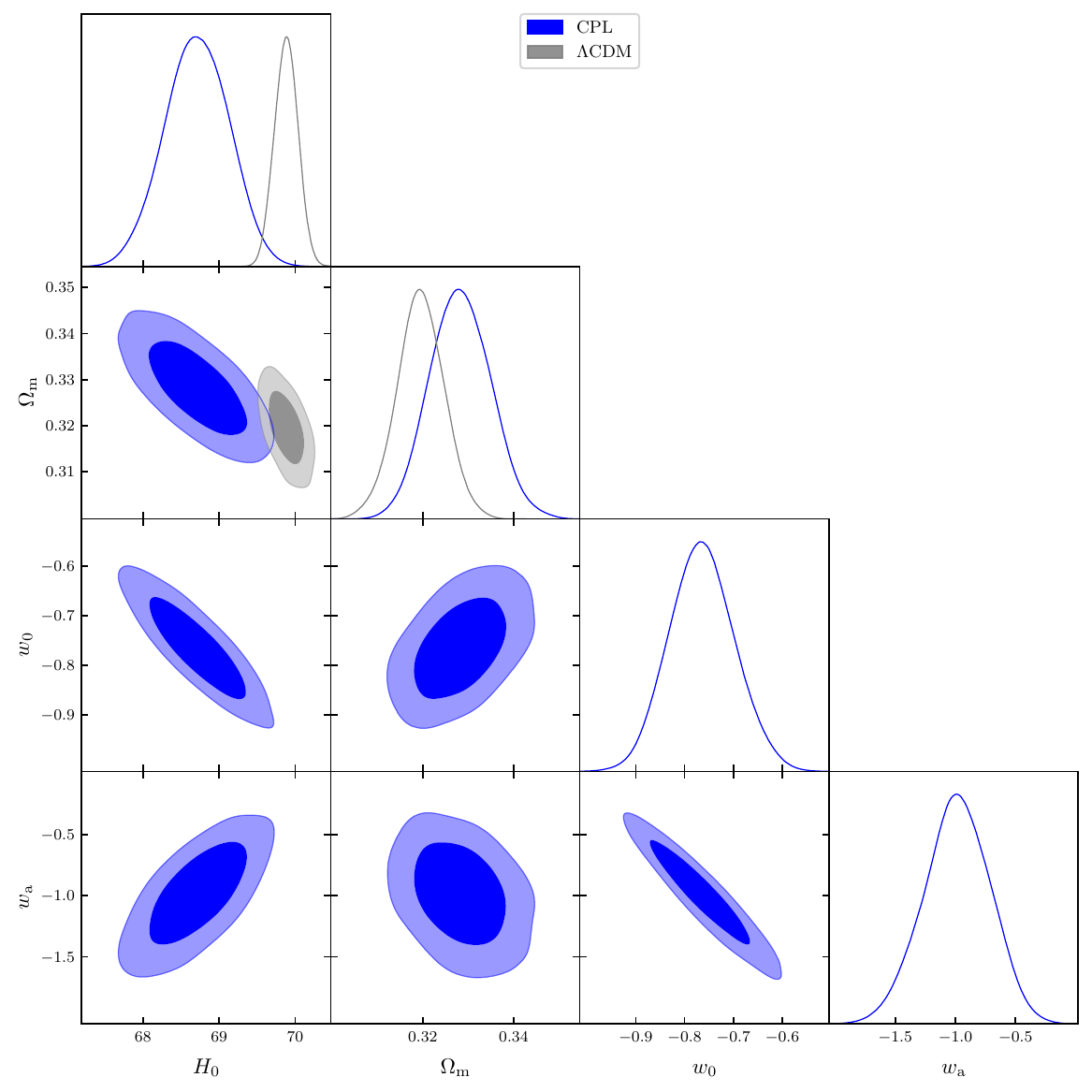}\includegraphics[width=0.5\linewidth]{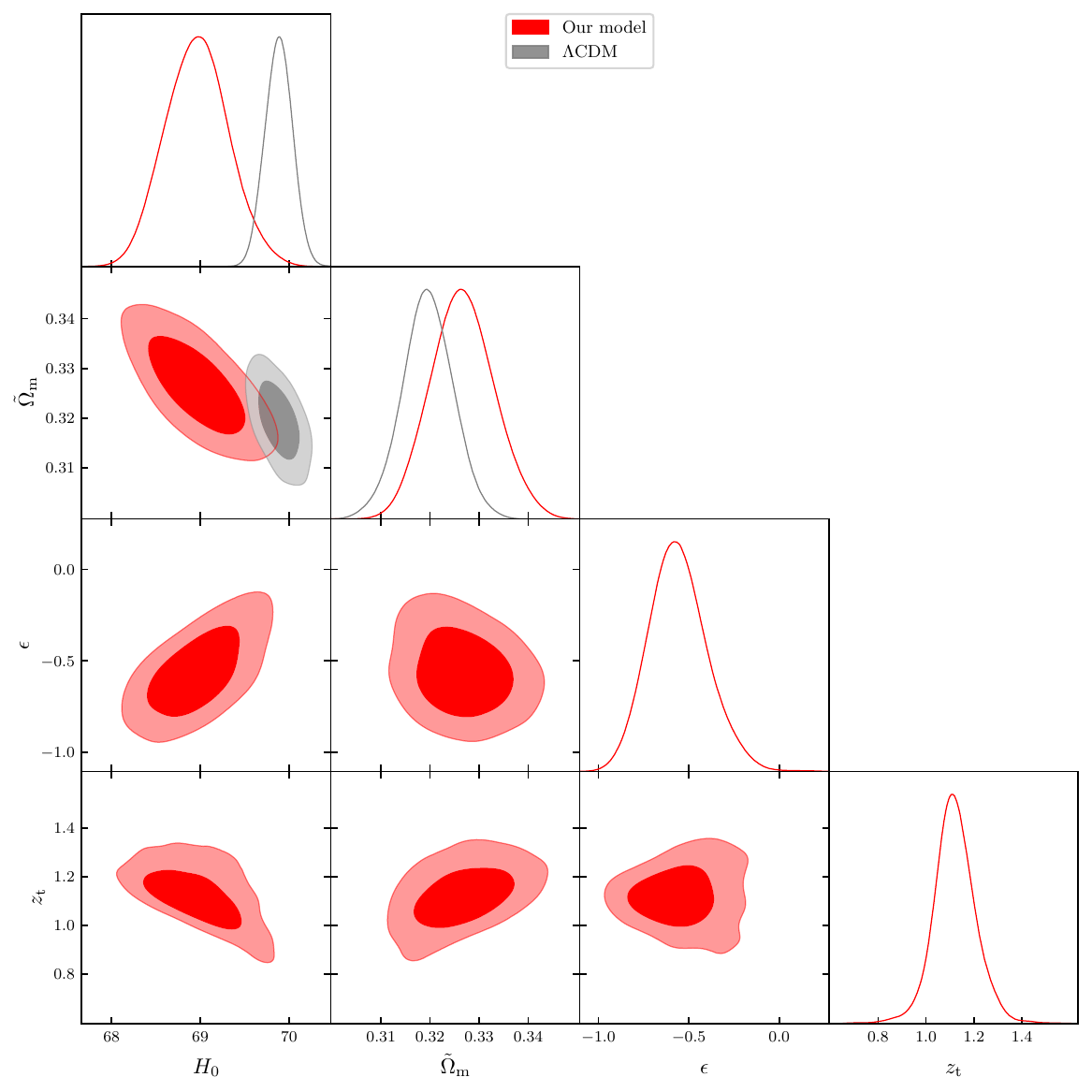}
    \caption{The posteriors from our MCMC explorations for the CPL parametrization (left) and the model proposed in this work (right). Interestingly, the marginalized posteriors on $H_0$ and $\Omega_m,\tilde{\Omega}_m$ are very similar, despite a very different phenomenology at intermediate redshifts.}
    \label{fig1}
\end{figure}
\section{Analysis}
Since our model modifies the expansion history only at late times, we consider the CMB distance priors developed in ~\citet{Chen:2018dbv} for the parameters $\theta_*, R$ and $w_b$. A caveat is in order: whilst both $\theta_*$ and $w_b$ can be considered model independent–the former measured directly in the CMB, and the latter strongly constrained by BBN– the prior on the shift parameter $R$ implies a few hypothesis on $\omega_c$ and pre-recombination physics, as discussed thoroughly in \citet{Pedrotti:2024kpn,Poulin:2024ken}. We also use the most recent BAO measurements from the DESI DR2 data release \citet{DESI:2025zgx}, in particular six measurements of distances $D_H$ and $D_M$ in the redshift range $0.5<z<2.5$. Finally, we consider the $\sim 1500$ Sn~Ia measurements in the DESY5 catalog \citet{DES:2024tys}. It has been argued by ~\citet{Efstathiou:2024xcq} that the evidence for dynamical dark energy reported in the DESI DR1 BAO paper ~\citet{DESI:2024mwx} might be due to systematics in the use of the DESY5 catalog. The DES collaboration, however, addressed these criticisms in \citep{DES:2025tir}. Recently, \citet{Efstathiou:2025tie} claimed that DESY5 data might be inconsistent with individual and combined measurements from DESI and Planck. In this paper we decide to use this catalog since it provides the strongest evidence in support of dynamical dark energy when combined with BAO measurements, and therefore allows to a fairer comparison of our model against the CPL parametrization.  

The main results of our MCMC exploration, comparing CPL and the parametrization in Eq.~\eqref{non-dust} with $\Lambda$CDM, are reported in Fig.~\ref{fig1}. We consider the MCMC chains converged
if the Gelman Rubin $R-1$ parameter \citep{Gelman:1992zz}, computed using GetDist \citep{Lewis:2019xzd}, is smaller than $R-1\leq 0.01$. Compared to $\Lambda$CDM, both the CPL parametrization and our model prefer lower values of $\Omega_m$, $\tilde{\Omega}_m$ and $H_0$ at present day. Notice that the parameter $\tilde{\Omega}_m$ on the right-hand side of Fig.~\ref{fig1} is not the present day matter density. As described in Eq.~\eqref{present-day-matter}, the two differ by a factor $\left(1+z_t\right)^{-\epsilon}\approx 1.5$ at the MAP values from Table \ref{Marginalized-constraints} ($\epsilon\approx -0.58$ and $z_t\approx 1.14$).

Table \ref{model:comparison} reports the values of the $\chi^2$ of the individual datasets computed at the maximum a posteriori (MAP), as well as the differences in the deviance information criteria (DIC) between the models and the $\Lambda$CDM \citep{Trotta:2008qt}. Interestingly, compared to CPL, our model performs slightly better in fitting the BAO data but slightly worse in fitting CMB and supernovae. We compute the preference for these models over $\Lambda$CDM using their respective p-values from the cumulative $\chi^2$ distribution, and find that both are preferred at more than $\geq 3 \sigma$ significance, but the two are essentially indistinguishable in the current analysis.  

\begin{figure}
    \centering
    \includegraphics[width=0.49\linewidth]{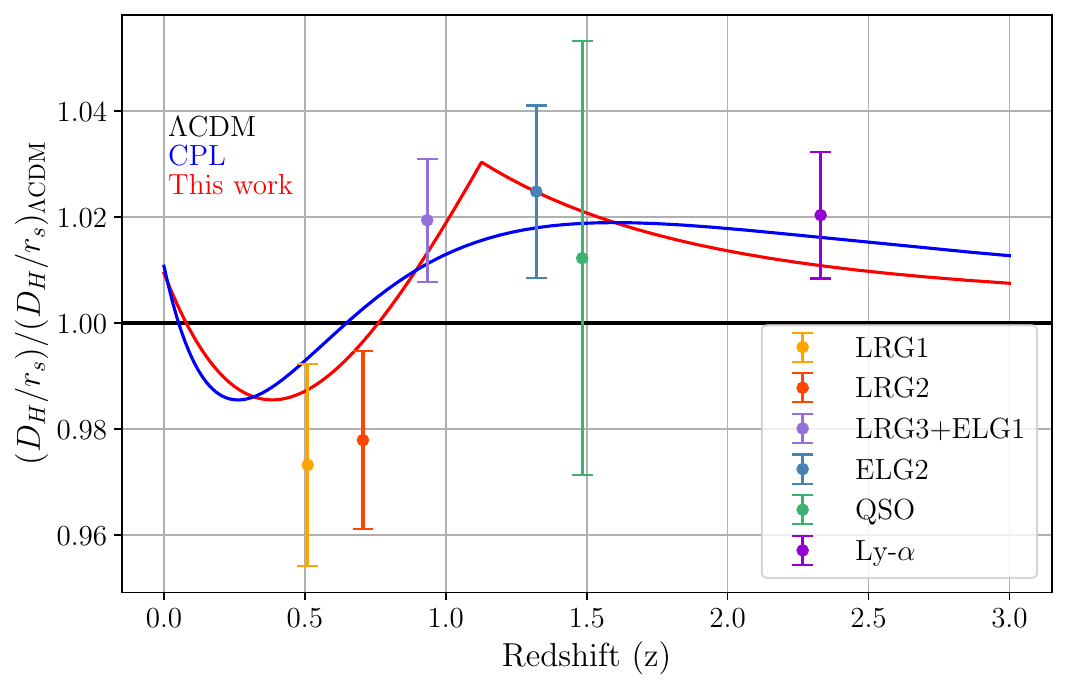}
    \includegraphics[width=0.49\linewidth]{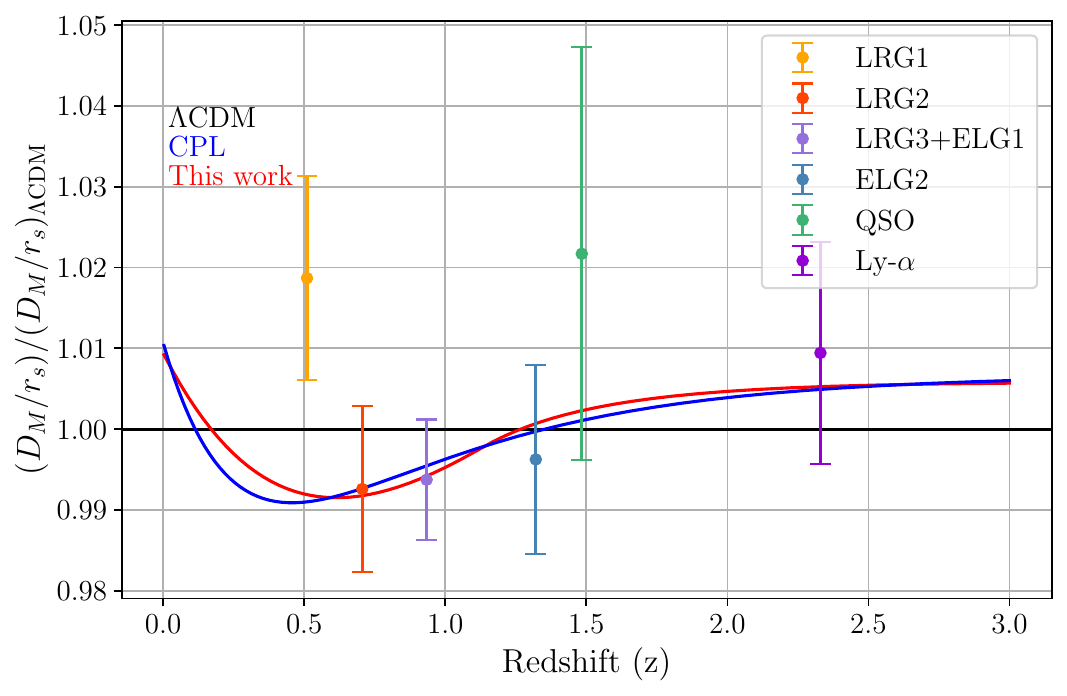}
    \caption{The behavior of the distances $D_H$ (left) and their integral $D_M$ (right), weighted using the fiducial $\Lambda$CDM best fit. As reflected by the variation of the corresponding $\chi^2$ distributions, our model offers a slightly better fit to the BAO data in our model compared to CPL. Notice that the error bars on the data points, due to the rescaling of the $\Lambda$CDM best fit, are larger than the original DESI data and misleadingly suggest that all the measurements are within $1-\sigma$ from $\Lambda$CDM. The sharp transition in our model at $z=z_t$ is reflected by the spike in the theoretical prediction for $D_H$.}
    \label{Distances}
\end{figure}

Fig.~\ref{Distances} shows the behavior of the BAO distances in the two models compared to the best fit of the fiducial $\Lambda$CDM model. One can see that while the integrated expansion history in CPL and our proposal is very similar (as reflected in the asymptotic behavior of $D_M$), differences in the actual $H(z)$ can be of the order of percent at redshifts $z \geq 0.5$, as reflected in the theoretical prediction for $D_H$. This implies that the two models, although rather degenerate at lower redshifts, can be distinguished in the future when more data at intermediate redshifts, such as those expected from Euclid, LSST and DESI will be available.

\begin{table}[h!]
\centering
\begin{tabular}{|c|cccccc|}
  \hline
  \multirow{2}{*}{Models} & \multicolumn{6}{c|}{Parameters} \\
                          & $\Omega_m$ (or $\tilde{\Omega}_m$) & $H_0$ & $w_0$ & $w_a$ & $\epsilon$ & $z_t$ \\
  \hline
  $\Lambda$CDM                 & $0.32 \pm 0.01$ & $69.89\pm0.15$ & -- & -- & -- & -- \\
  \hline
  CPL                    & $0.33 \pm 0.01$ & $68.70 \pm 0.42$ & $-0.76 \pm 0.07$ & $-0.97 \pm 0.28$ & -- & -- \\
  \hline
  This Work                    & $0.33 \pm 0.01$ & $68.94 \pm 0.35$ & -- & -- & $-0.58 \pm 0.16$ & $1.14 \pm 0.08$ \\
  \hline
\end{tabular}
\caption{Results of our MCMC exploration for the marginalized constraints on the cosmological parameters of the different models. Notice that the $\Omega_m$ relative to the model proposed in this work is actually $\tilde{\Omega}_m$, as described by Eq.~\eqref{present-day-matter}.}
\label{Marginalized-constraints}
\end{table}

\begin{table}
\centering
\begin{tabular}{lcccccc}
\hline\hline
\multirow{2}{*}{Model} & \multicolumn{3}{c}{$\chi^2_{\rm MAP}$} & \multirow{2}{*}{$\Delta\chi^2_{\rm MAP}$} & \multirow{2}{*}{$\Delta$(DIC)} & \multirow{2}{*}{Significance} \\
                       & BAO        & SN Ia        & CMB        &                                           &                                &                               \\ \hline
$\Lambda$CDM           & 17.17       & 1644.08       & 0.75       & --                                        & --                             & --                            \\
CPL                    & 9.19       & 1638.09       & 0.53       & -14.19                                    & -11.18                         & 3.41$\sigma$                  \\
This work              & 7.90       & 1639.71       & 0.63       & -13.76                                    & -9.57                          & 3.31$\sigma$                  \\ \hline\hline
\end{tabular}\\
\caption{The $\chi^2$ distributions of the individual datasets for the best fit (MPA) values of each model, and the differences in their DIC values with respect to $\Lambda$CDM. }
\end{table}\label{model:comparison}


\section{Interpretation and Discussion}

As already mentioned, our parametrization finds a natural interpretation in the context of cosmological backreaction. Indeed, one would naively expect a transition in the dark matter equation of state parameter once its fluid elements get rearranged in the complex network of large scale structures in the cosmic web. A posteriori, this speculation is supported by data preferring a transition redshift $z_t\approx \mathcal{O}(1)$, roughly at the time when density perturbations on the Mpc scale become non-linear. This feature somehow resembles the coincidence problem in the context of dynamical dark energy. Indeed, standard CPL seems to suggest a shift in the dark energy behaviour around redshift $z\sim0.5$, roughly at the transition from dark matter to dark energy domination.In our model the transition happens at the redshift where the variance of the matter density on scales of $\leq5$ Mpc is of order 1.

We stress indeed that the model recovers the $\Lambda$CDM limit when $\varepsilon=0$ or $z_t=0$, but this region of the parameter space seems excluded at more than $2 \sigma's$ (see Fig.~\ref{fig1}).

There is of course an elephant in the room asking which type of non-linear effects could give rise to such a negative $\epsilon$ as the one preferred by the data. Indeed, previous investigations on the subject of cosmological backreaction from non-linearities, see for example \citet{Baumann:2010tm}, excluded the possibility that matter non-linearities can produce an effective negative equation of state parameter $w$. On the other hand, other works have suggested that backreaction from the cosmic web can effectively mimic the appearance of spatial curvature, see for example Refs.~ \citep{Bolejko:2017lai,Heinesen:2020sre}, whose equation of state parameter can be negative. ~\citet{Giani:2024nnv} concluded that a negative effective $w$ can be expected if large voids of radius $1-10$ Mpc are the driving force of backreaction effects in the cosmic web at late times. 
In light of the above investigations, it seems more likely that large underdense inhomogeneities at low redshift could contribute to an $\epsilon$ of order $\sim -0.5$, rather than local inhomogeneities such as an LTB void or large scale anisotropies of Bianchi type, given the currently strong constraints on both these scenarios \cite{Camarena:2021mjr,Akarsu:2019pwn}

\begin{figure}[h!]
    \centering
    \includegraphics[width=1\linewidth]{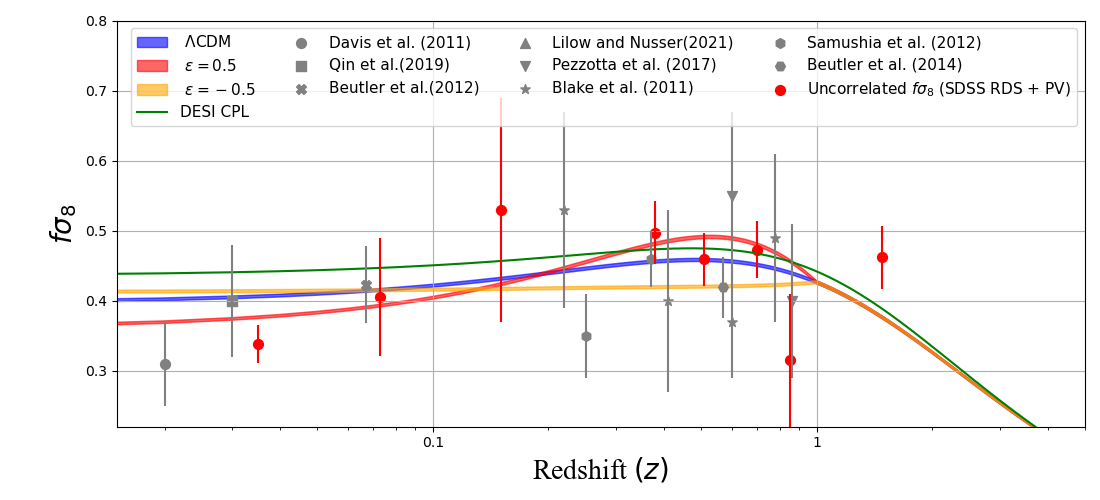}
    \caption{Predictions for the evolution of $f\sigma_8$ for the $\Lambda$CDM (blue) and our model, fixing $\Omega_m=\tilde{\Omega}_m=0.3$, $H_0=70$ km/s/Mpc and a transition at redshift $z_t=1$ with $\epsilon = \pm 1/2$ (red and yellow). Red dots indicate a number of recent, uncorrelated $f\sigma_8$ measurements from \citet{eBOSS:2020yzd,Lai:2022sgp,Said:2020epb}, and the grey points a collection of correlated measurements usually considered in the literature (from \citet{Davis:2010sw,Qin:2019axr,BOSS:2013uda,Lilow:2021omg,Pezzotta:2016gbo,2011MNRAS.415.2876B,BOSS:2013yzh,2012MNRAS.423.3430B})}
    \label{fsigma8}
\end{figure}

Despite the above theoretical issues, it is remarkable that a simple parametrization as the one in Eq.~\eqref{non-dust} for the matter fluid evolution can result in such an interesting phenomenology and be competitive with the CPL parametrization. Furthermore, if we assume that the backreaction interpretation is correct, it allows for a physically motivated prediction for the behavior of linear cosmological perturbations. Indeed, as discussed in \citep{Giani:2024nnv} the non-linearities are expected to impact the background evolution but not the linear interactions between dust fluid elements. Therefore, the only modifications to the evolution equation for the density contrast on sub-horizon scales
\begin{equation}
    \delta'' + 2H\delta -4\pi G_N\rho_m=0\;,
\end{equation}
are due to the different behavior of $H$ and $\rho_m$. As a consequence, the predictions for the growth rate of structures are easily distinguishable. Fig.~\ref{fsigma8} compares the evolution of $f\sigma_8$ predicted by Eq.~\eqref{non-dust} and the $\Lambda$CDM one assuming $\Omega_m= \tilde{\Omega}_m = 0.3$ and $H_0 = 70$ km/s/Mpc. For the CPL parameterization, our ignorance about the perturbative nature of the underlying dark energy fluid prevents us from analyzing similar observables unless we make some additional assumptions. We report in Fig.~\ref{fsigma8} the behaviour of $f\sigma_8$ for the $w_0w_a$ model from CLASS, assuming that dynamical dark energy is not clustering, for the DESI best fit.

In the past few months, while this work has been developed, some works have studied the potential of a modified matter evolution rather than dynamical dark energy \citep{Kumar:2025etf,Yang:2025ume}, and a few days before the submission of this paper \citet{Chen:2025wwn} considered a very similar step function model, but without providing a similar and thorough Bayesian analysis (our predicted phenomenology is, however, in agreement). Further analyses of non-standard matter evolution were considered in \citep{Wang:2025hlh,Yao:2025kuz,Li:2025dwz,Braglia:2025gdo}.

These alternative descriptions are probably driven
by the desire to avoid the issues of phantom crossing implied by the use of CPL parametrization and the disagreement between the inferred value of $\Omega_m$ from different datasets. In these regards, the parametrization advanced in this work possesses both these features in a rather economical way: there is no phantom crossing, and it is expected to infer a different extrapolated matter density at different times. Furthermore, the model proposed here and $w_0w_a$ have rather different behavior at intermediate redshift and we expect that future data from for Euclid, LSST and DESI will be able to distinguish between the two.

There is one further advantage of our parametrization worth discussing. As mentioned earlier (and shown explicitly in the left panel of Fig.~\ref{fig1}), the lower values for $\Omega_m$ obtained using the CPL parametrization seem to worsen the Hubble tension. Our results show that in fact the model prefers a lower matter density at earlier times $\tilde{\Omega}_m$ than in the $\Lambda$CDM model. However, the presence of a negative $\epsilon$ makes it dilute slower at late times, so the density of physical matter is greater than a factor $\sim 1.5$. At the very least, the phenomenology presented in this work might provide a framework to deal with the $\Omega_m$ and $H_0$ tensions without incurring in the proverbial \textit{short blanket} dilemma. In this work, we focus on the comparison between our model and the CPL parametrization, without including local $H_0$ information, and therefore it is not surprising to find values of the Hubble constant compatible with Planck. We will investigate how this picture changes including datasets relevant for the Hubble tension and $f\sigma_8$ measurements, such as full-shape measurements, low redshift Sn~Ia and fundamental plane in an upcoming paper.

\section{Acknowledgements}
We are grateful to Sunny Vagnozzi for useful comments and suggestions. LG acknowledges support from the Australian Government through the Australian Research Council Centre of Excellence for Gravitational
Wave Discovery (OzGrav). RvM is suported by Funda\c{c}\~ao de Amparo \`a Pesquisa do Estado da Bahia (FAPESB) grant TO APP0039/2023.
\bibliography{sample7.bib}{}
\bibliographystyle{apsrev4-1}



\end{document}